\documentclass{elsarticle} 
\graphicspath{{pics/}}
\usepackage{amsmath}
\usepackage{amsfonts}
\usepackage{graphicx}
\usepackage{color}

\begin{document}

\begin{frontmatter}

\title{Magnetic field generation by intermittent convection}
\author[systec,samara]{R. Chertovskih\corref{mycorrespondingauthor}}
\cortext[mycorrespondingauthor]{Corresponding author}
\ead{roman@ssau.ru}

\author[ita,wiser]{E.L. Rempel} 
\author[ita]{E.V. Chimanski}

\address[systec]{Research Center for Systems and Technologies, Faculty of Engineering, 
\\University of Porto, Rua Dr. Roberto Frias, s/n, Porto 4200--465, Portugal}
\address[samara]{Samara National Research University, Moskovskoye Shosse 34,\\ Samara 443086, Russian
Federation}
\address[ita]{Aeronautics Institute of Technology, S\~ao Jos\'e dos Campos, S\~ao Paulo 12228--900, Brazil}
\address[wiser]{National Institute for Space Research and World Institute 
for Space Environment Research, S\~ao Jos\'e dos Campos, S\~ao Paulo 12227--010, Brazil}

\begin{keyword}
Magnetohydrodynamics\sep Rayleigh-B\'enard convection\sep  convective dynamo\sep 
intermittency.
\end{keyword}

\begin{abstract}
Magnetic field generation in three-dimensional Rayleigh-B\'enard convection
of an electrically conducting fluid is studied numerically 
by fixing the Prandtl number at $P=0.3$ and varying the Rayleigh number (Ra) as a control parameter. A
recently reported route to hyperchaos involving quasiperiodic regimes, crises and
chaotic intermittent attractors is followed, and the critical magnetic Prandtl number
($P_m^c$) for dynamo action is determined as a function of Ra. A mechanism
for the onset of intermittency in the magnetic energy is described, the most beneficial
convective regimes for dynamo action in this transition to weak turbulence are identified, 
and the impact of intermittency on the dependence
of $P_m^c$ on Ra is discussed.
\end{abstract}

\end{frontmatter}

\section{Introduction} 
Thermal convection plays an important role in magnetic field generation 
in planets and stars.
For instance, the Earth's magnetic field is sustained by a dynamo
process driven by convective flows in the liquid iron outer core, whereby
thermal energy is transformed into kinetic energy and then
into magnetic field energy.  The fluid motions are
driven by buoyancy forces and are strongly affected by the Lorentz force, due to
the strength of the Earth's magnetic field in the core, as well as the Coriolis
force, due to the Earth's rotation \cite{zhang99}. Historical data on
geomagnetic field evolution based on the paleomagnetic record that can form in
rocks reveal nonperiodic intermittent time series, both for the field intensity
and for the frequency of polarity reversals \cite{resh16}. By intermittent, we mean that
the field undergoes recurrent and aperiodic switching between two 
qualitatively different states. Thus,
the geomagnetic field randomly alternates between stronger (``bursty'') and weaker (``laminar'') phases. 
The relation between the field intensity and the frequency
of reversals is still not clear. For instance, Pr\'evot {\it et al}. \cite{prevot90} conjectured that the
Mesozoic dipole low corresponded roughly to a progressive decrease in the
average frequency of reversals, which contrasts with more recent data that
indicate that the time-averaged field has been higher during periods without
reversals in the past two million years, whereas more reversals are expected
during periods of weak field intensity \cite{valet05}. In fact, there may be no
simple correlation between reversal rate and intensity~\cite{heller02},
but overall the geodynamo seems to exhibit a strongly chaotic/intermittent rather than
periodic behavior. Intermittency is also observed in the solar magnetic 
activity as recorded by the grand minima of the sunspots cycle \cite{beer98}.

Several numerical experiments have tried to capture the chaotic intermittent
features of the geomagnetic and solar dynamos. The main difficulty in such studies is that a
dynamo mechanism cannot sustain an axisymmetric magnetic field, due to Cowling's
anti-dynamo theorem \cite{cowling33}, therefore, fully three-dimensional models have to
be employed for a self-consistent simulation. Both global (spherical)
\cite{glatz99,ol09,ray16} or local (planar) \cite{physd,calk15} 
three-dimensional convective models have been used. However, parameter values
corresponding to the planetary and stellar interiors require a huge amount of computations which
cannot be performed even on modern high performance computers in a reasonable time.

A set of dimensionless parameters has been defined to describe convective
dynamos.  Some of them are the Rayleigh number (Ra), that measures the
magnitude of the thermal buoyancy force, the Prandtl number ($P$), which is the
ratio of kinematic viscosity to thermal diffusivity, and the magnetic Prandtl
number ($P_m$), defined as the ratio between kinematic viscosity and magnetic
diffusivity.  One of the main problems in dynamo theory is to determine the
relation between the values of those parameters and the onset of dynamo action.
For a review of scaling properties of convective--driven dynamo models in
rotating spherical shells, see \cite{chris06}.  Here, we are
interested in magnetic field generation in Rayleigh-B\'enard convection (RBC) for
moderately low Prandtl number (the Prandtl number in the Earth's outer core has
been estimated to be between 0.1 and 0.5 \cite{olson07, fearn}).  
In this context, strong dependence of the magnetic fields generated by
convective flows on the value of the Prandtl number was found for $0.2 \leq P
\leq 5$ \cite{soltis14}. Analysis of the kinematic dynamo problem by
Podvigina \cite{podv08} showed that convective attractors for $P = 0.3$ favour
the magnetic field generation in comparison to the attractors found for larger values of $P$
($P = 1$ and $P=6.8$). In recent multiscale analysis of the RBC 
dynamo problem in the presence of rotation \cite{calk16}, 
low Prandtl number convection was also found to be beneficial for magnetic field generation. 
In the present work, we follow our recently published analysis of
transition to chaos and hyperchaos 
in RBC as a function of the Rayleigh number
for $P=0.3$ \cite{epl}. 
Chaotic systems are characterized by aperiodic motions 
that are sensitive to initial conditions, i.e., close initial conditions tend
to locally diverge exponentially with time, with the mean divergence rate measured by
a positive maximum Lyapunov exponent. In hyperchaos a system presents more than 
one positive Lyapunov exponent, implying that divergence occurs in more than
one direction among a set of orthogonal directions in the phase space. 
As the number of positive exponents increases, the resulting dynamics becomes
more irregular in space and time, thus, hyperchaotic systems have been termed 
weakly turbulent \cite{manneville}.
We assume the fluid to be electrically conducting 
and add an initial seed magnetic field to investigate the onset of dynamo. 
The intermittency in the velocity field described in ref. \cite{epl} as being due to global
bifurcations constitutes the starting point for our search for an explanation for
the intermittent dynamo. We stress that our hydrodynamic regimes are in a transition
state between weak chaos and turbulence, therefore, the system displays a complex switching
between highly different phases that is typically very difficult to be studied, since 
the convergence of average quantities is much 
slower than in fully developed turbulence, and the meaning of those averages may be rather deceptive
if they do not take into account information
about the different phases of the flow.

Usually, large Rayleigh numbers are beneficial for magnetic field generation in
spherical shells simulations \cite{busse00}, {\it i.e.}, the critical $P_m$ for dynamo
action decreases with increasing Ra.  For moderate values of Ra, in Calkins {\it et al}.
\cite{calk16} a similar dependence in plane layer dynamos was found, while in refs.
\cite{podv08,podv06} it is reported  that for Ra beyond a certain threshold the behaviour
of the critical $P_m$ ceases to be monotonic. In this work we describe a scenario
where the non-monotonic behaviour of the critical $P_m$ is explained based on an
analysis of the intermittency in the convective attractors. Thus, our goals are
threefold: $i$) to explain a mechanism for intermittency in magnetic field
fluctuations in RBC; $ii$) to describe how this type of intermittency is
responsible for the non-monotonic behaviour of the critical $P_m$ as a function of Ra
and $iii$) to detect what are the best hydrodynamic regimes for magnetic field
amplification, that is, the regimes where dynamo action takes place for smaller
$P_m$, among the hydrodynamic attractors in the RBC system in transition to
turbulence.

\section{Statement of the problem and solution} 

We consider a newtonian incompressible fluid flow in a horizontal plane layer; 
the fluid is electrically conducting and uniformly heated from below and cooled from above. 
Fluid flow is buoyancy-driven and the Boussinesq approximations 
(see, {\it e.g.}, \cite{chandra}) are assumed. 
In a Cartesian reference frame with the
orthonormal basis $({\bf e}_1, {\bf e}_2, {\bf e}_3)$, where ${\bf e}_3$
is opposite to the direction of gravity, the equations governing the
magnetohydrodynamic (MHD) system are \cite{meneguz89}:
\begin{gather}
\frac{\partial {\bf v}}{\partial t}=
P\nabla^2{\bf v}
+{\bf v}\times(\nabla \times {\bf v})\nonumber
-{\bf b}\times(\nabla \times {\bf b})\\
+P{\rm Ra}\theta{\bf e}_3-\nabla p,
\label{NSlayer} \\
\frac{\partial{\bf b}}{\partial t}=
\frac{P}{P_m}\nabla^2{\bf b}+
\nabla\times({\bf v}\times{\bf b}),
\label{Blayer} \\
\frac{\partial\theta}{\partial t}=
\nabla^2\theta
-({\bf v}\cdot\nabla)\theta+v_3,
\label{Tlayer} \\
\nabla\cdot{\bf v}=0,
\label{Vsol}\\
\nabla\cdot{\bf b}=0,
\label{Bsol}
\end{gather}
where ${\bf v}({\bf x},t)=(v_1, v_2, v_3)$ is the fluid velocity, 
${\bf b}({\bf x},t)=(b_1, b_2, b_3)$ the magnetic field,
$p({\bf x},t)$ the pressure, $\theta({\bf x},t)$ is the difference between
the flow temperature and the linear temperature profile;
the spatial coordinates 
are ${\bf x}=(x_1,x_2,x_3)$ and $t$ stands for time. 
The non-di\-men\-si\-onal parameters are the Prandtl number, 
$P=\nu/\kappa$, the magnetic Prandtl number, $P_m=\nu/\eta$, and the Rayleigh number, 
${\rm Ra}=\alpha g \delta T d^3/(\nu \kappa)$, where $\nu$ is the kinematic viscosity,
$\kappa$ the thermal diffusivity, $\eta$ the magnetic diffusivity,    
$\alpha$ the thermal expansion coefficient, $g$ the gravity acceleration, 
$\delta T$ the temperature difference between the layer boundaries,
and $d$ the vertical size of the layer.
The units of length and time are $d$ and the vertical heat 
diffusion time, $d^2/\kappa$, respectively;
${\bf v},$ ${\bf b}$ and $\theta$ are measured in units of 
$\kappa/d,$ $\sqrt{\mu_0 \rho} \kappa/d$ and $\delta T$,
respectively. Here  $\mu_0$ stands for the vacuum magnetic 
permeability and $\rho$ the mass density.

The horizontal boundaries of the plane layer, $x_3=0$ and ${x_3=1}$, 
are assumed to be stress-free,
$\partial v_1/\partial x_3=\partial v_2/\partial x_3=v_3=0$,
electrically perfectly conducting, 
$\partial b_1/\partial x_3=\partial b_2/\partial x_3=b_3=0$, and
maintained at constant temperatures, $\theta=0$. A square convective
cell is considered, ${\bf x}\in[0,L]^2\times[0,1]$, all the fields are
periodic in the horizontal directions, $x_1$ and $x_2$, with period $L$. 

We study magnetic field generation by the hyperchaotic convective 
attractors found in \cite{epl} for $L=4$, \mbox{$P=0.3$} and 
$2160\le{\rm Ra}\le5000$ (the critical Rayleigh number 
for the onset of convection is $\rm Ra_c$=657.5 \cite{chandra}). 
For a given value of Ra, 
the initial condition for the hydrodynamic part of the system, ${\bf v}({\bf x},0)$ and 
$\theta({\bf x},0)$, is taken from the corresponding convective attractor; the initial magnetic field is 
${\bf b}({\bf x},0)=(\cos(\pi x_2/2), 0, 0)$ scaled such that the magnetic energy 
$E_b(0)=10^{-7}.$ For each convective attractor the governing equations (1)--(5) are integrated forward in
time for a certain value of $P_m$ in order to estimate the critical magnetic 
Prandtl number, $P_m^c$. Thus, if $P_m>P_m^c$ the magnetic field is maintained 
in the system (dynamo), for  $P_m<P_m^c$, the magnetic field decays (no dynamo). 
We consider values of $P_m$ from 1 to 10 with step 1; the value is regarded to be 
subcritical if $E_b(t)<10^{-50}$ for a time interval greater than 500,
otherwise it is supercritical and the generated magnetic field is analysed. 

For a given initial condition, equations (1)--(5) are
integrated numerically using the standard pseudospectral
method \cite{canuto1}: the fields are represented as Fourier series in
all spatial variables (exponentials in the horizontal directions,
sine/cosine in the vertical direction), derivatives are computed in the
Fourier space, multiplications are performed in the physical space, and
the Orszag 2/3-rule is applied for dealiasing. 
The system of ordinary differential equations for
the Fourier coefficients is solved using the third-order exponential
time-differencing method ETDRK3 \cite{coxmatt} with constant step
$5\cdot10^{-4}$. At each time step the fields 
$\bf v$ and $\bf b$ are projected onto space of solenoidal functions
by solving the Poisson equation in the Fourier space.
Spatial resolution is
$86\times86\times43$ Fourier harmonics (multiplications were performed
on a uniform $128\times128\times64$ grid). 
For all solutions, time-averaged
energy spectra of the magnetic and velocity fields decay as a function of the 
wavenumber at least by 3 and 7 orders of magnitude,
respectively. Several simulations with the doubled resolution showed no significant 
change in the results.
In computations we traced the kinetic, $E_v(t)$, and magnetic, $E_b(t)$, energies, 
which are the $L_2$ norm of the squared field normalized by the volume of the convective cell.
We also computed the critical magnetic Reynolds number, ${\rm Re_m^c}=P_m^c L \sqrt{<E_v>}/P,$ 
where angle brackets stand for time-averaging.

\begin{figure}
\centerline{\includegraphics[scale=1.0]{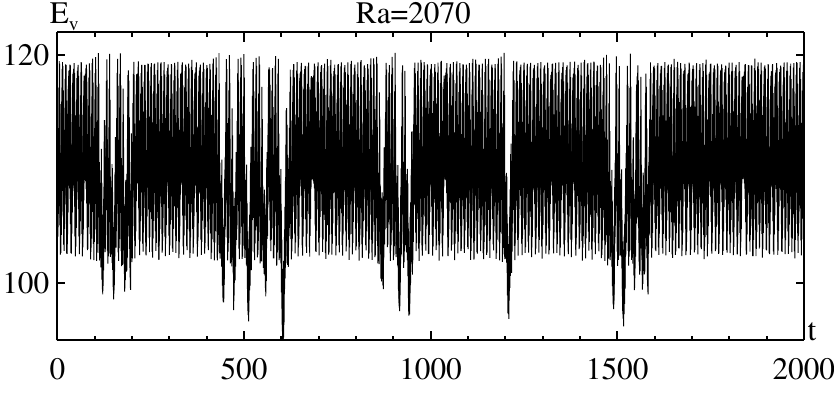}}
\vspace*{-0.8cm}
\centerline{\hspace*{-8.2cm}(a)}
\vspace*{0.30cm}
\centerline{\includegraphics[scale=1.0]{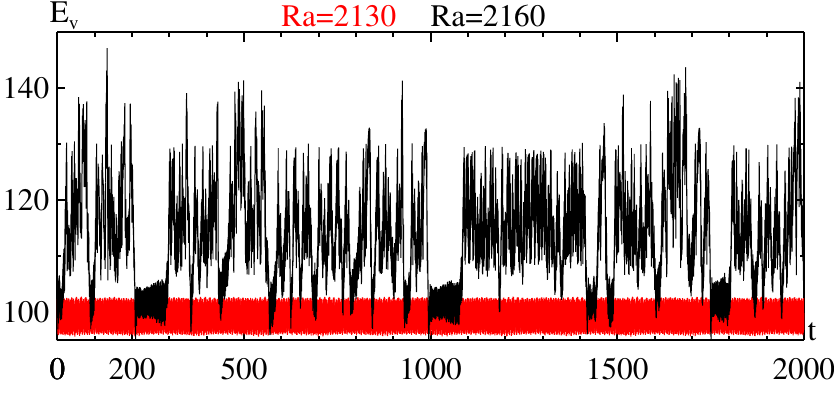}}
\vspace*{-0.8cm}
\centerline{\hspace*{-8.2cm}(b)}
\vspace*{0.30cm}
\centerline{\includegraphics[scale=1.0]{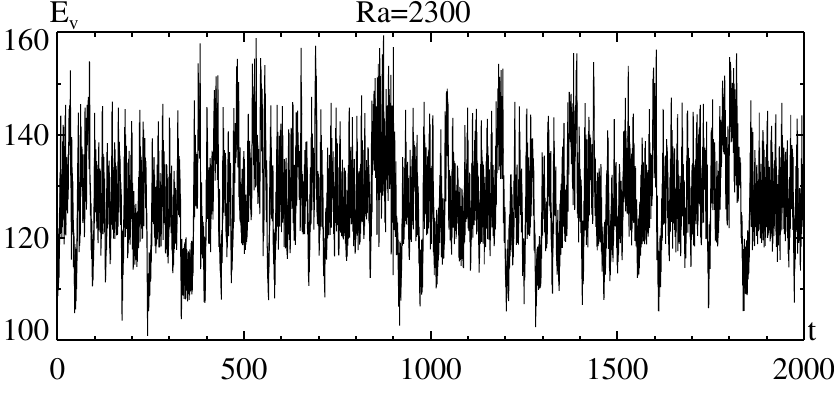}}
\vspace*{-0.6cm}
\centerline{\hspace*{-8.2cm}(c)}
\vspace*{0.10cm}
\centerline{\includegraphics[scale=1.0]{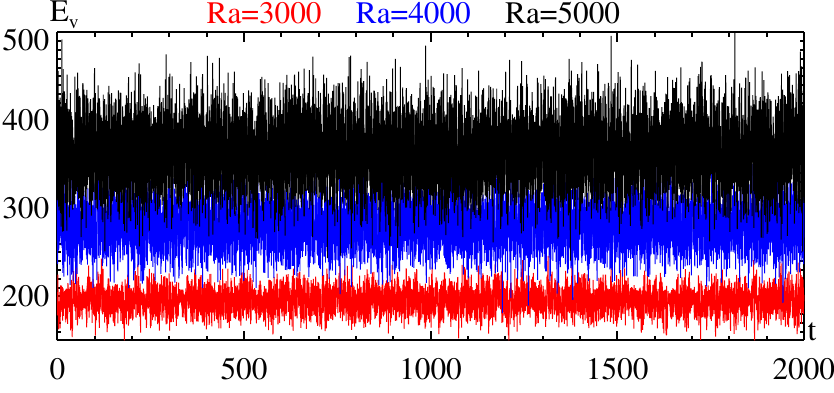}}
\vspace*{-0.6cm}
\centerline{\hspace*{-8.2cm}(d)}
\caption{(Color online) Evolution of the kinetic energy in the absence of magnetic field for 
the attractors of the convective system at Ra=2070 (a), 2130, 2160 (b),  
2300 (c) 3000, 4000 and 5000~(d).
}
\label{fig:conven}
\end{figure}

\section{Results} 

The present work studies magnetic field
generation by the branch of hyperchaotic attractors of the convective system for
$\rm{Ra}\ge2140$ \cite{epl}. This branch is formed in an interior crisis,
whereby a quasiperiodic attractor collides with a background nonattracting chaotic
set (a chaotic saddle) to form an enlarged chaotic attractor, where trajectories intermittently switch
between the former quasiperiodic attractor and the surrounding chaotic saddle. In what
follows we briefly describe the convective regimes in the absence of magnetic field 
in this interval of Ra, also mentioning the pre-crisis attractors at Ra=2070 and Ra=2130 
(see \cite{epl} for more details); the considered values of Ra and the temporal 
behaviour of the corresponding hydrodynamic attractors are summarized in table~\ref{tab:pmc}.

\subsection{Convective states}

\begin{table}[t!]
\caption{Types of the convective attractors considered in the paper, 
ratio of the Rayileigh number to its critical value at the onset of convection;
the critical values of the magnetic Prandtl number, $P_m^c,$
and the corresponding critical magnetic Reynolds number, ${\rm Re_m^c},$ 
computed for the middle value of $P_m^c$ in the corresponding interval. 
\label{tab:pmc}}
\begin{center}
\begin{tabular}{|ccccc|}
\hline
Ra  &Ra/$\rm Ra_c$& Type &$P_m^c$    & ${\rm Re_m^c}$\\\hline
2070&3.1&chaotic      &  $9<P_m^c<10$&1320 \\\hline
2130&3.2&quasiperiodic&  $2<P_m^c<3$ &332  \\\hline
2160&3.3&hyperchaotic&  $6<P_m^c<7$  &922  \\\hline
2300&3.5&hyperchaotic&  $7<P_m^c<8$  &1129 \\\hline
3000&4.6&hyperchaotic&  $8<P_m^c<9$  &1584 \\\hline
4000&6.1&hyperchaotic&  $6<P_m^c<7$  &1453 \\\hline
5000&7.6&hyperchaotic&  $5<P_m^c<6$  &1399 \\\hline
6000&9.1&hyperchaotic&  $4<P_m^c<5$  &1269 \\\hline
\end{tabular}
\end{center}
\end{table}

In the absence of magnetic field, the convective regime for Ra=2070 is chaotic 
(see fig.~\ref{fig:conven}(a)).
This convective state loses its stability for ${\rm Ra}>2070$, originating the chaotic 
saddle mentioned above. 
For Ra=2130, the sole attractor in the convective system is quasiperiodic with three incommensurate time frequencies.
The evolution of its kinetic energy is shown in fig.~\ref{fig:conven}(b)(red line) and in what follows it is 
referred to as the regular (convective) state. For Ra$\ge2140$, its stability is lost, 
and a new hyperchaotic attractor rises as the sole attractor of the convective system.
Its intermittent nature near its birth is revealed by the kinetic-energy time series (see
fig.~\ref{fig:conven}(b)(black line) for the attractor at Ra=2160). The trajectory in the phase
space visits the destabilized regular state at, {\it e.g.}, $200\le t\le 300$
and $1000\le t\le 1100$, as well as bursty chaotic phases related to the destabilized chaotic
attractor (compare the energy levels of the bursty phases of figs.~\ref{fig:conven}(b)(black line)
and ~\ref{fig:conven}(a)). The average time spent near the regular state is
shortened for increasing Ra (cf. Ra=2160 in fig.~\ref{fig:conven}(b) and Ra=2300 in fig.~\ref{fig:conven}(c);
see also fig.~6 in \cite{epl}). For ${\rm Ra}\ge3000$ the laminar phases in the 
intermittency are no longer recognizable at the time scale shown
(see the kinetic energy evolution for 
attractors at Ra=3000, 4000 and 5000 in fig~\ref{fig:conven}(d)). 
The absence of regular phases is to be expected, since those phases
are shortened as one moves away from the interior crisis point at ${\rm Ra}\approx 2140$~\cite{romeiras87}.

\subsection{Magnetic field generation}

\begin{figure}[!]
\centerline{\includegraphics[scale=1.0]{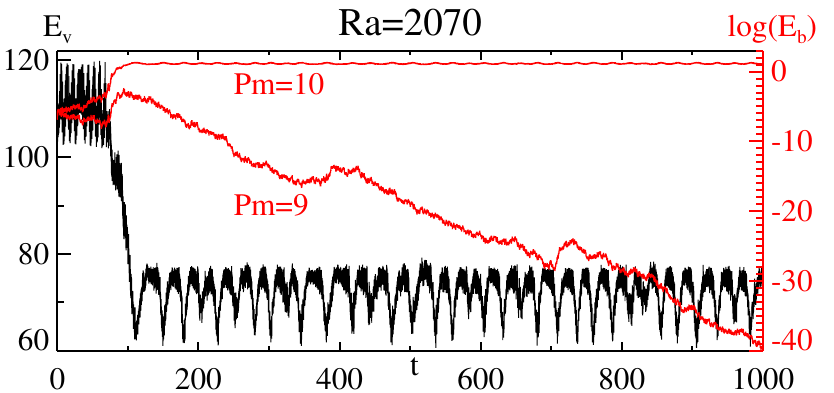}}
\vspace*{-0.5cm}
\centerline{\hspace*{-8.2cm}(a)}
\centerline{\includegraphics[scale=1.0]{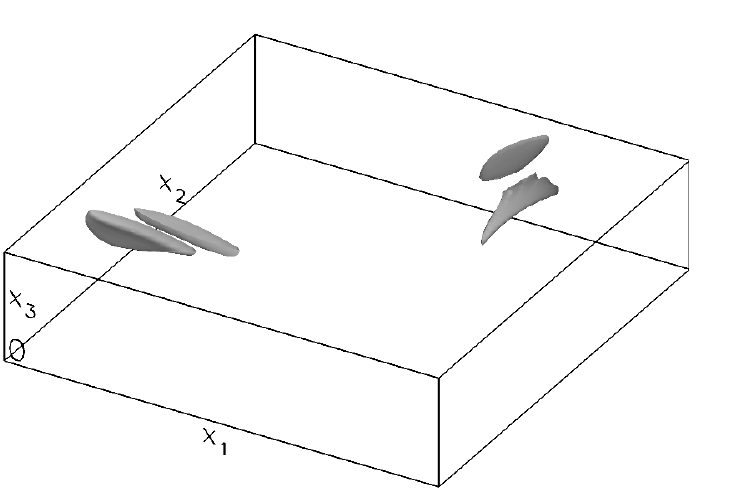}}
\vspace*{-0.5cm}
\centerline{\hspace*{-8.2cm}(b)}
\caption{(Color online) 
(a): Evolution of the kinetic (black line, left axis) and logarithm of the magnetic
(red line, right axis, non-decaying in time) energies for the convective attractor at Ra=2070 
and $P_m=10$. Magnetic energy decay for the same Ra and $P_m=9$ is
represented by the lower red line. (b): Snapshot at $t=400$ of isosurfaces of magnetic energy density,
at a level of a third of the maximum, for the MHD state at Ra=2070 and $P_m=10$ shown in (a); 
one periodicity cell is displayed.
}
\label{fig:ra2070pm10}
\end{figure}

\begin{figure}[!]
\centerline{\includegraphics[scale=1.0]{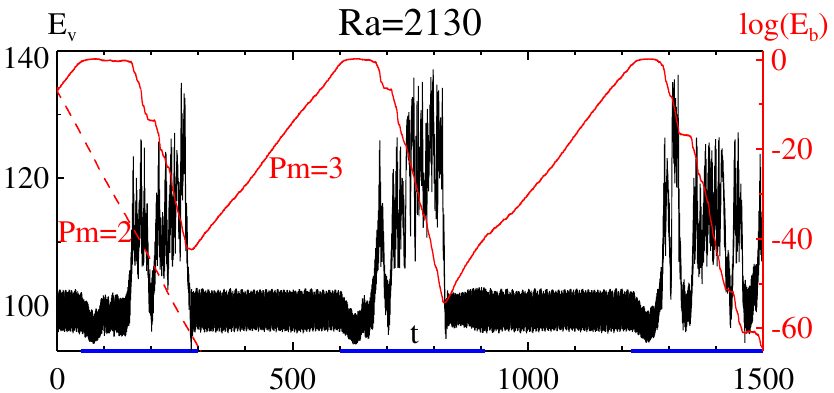}}
\vspace*{-0.5cm}
\centerline{\hspace*{-8.2cm}(a)}
\centerline{\includegraphics[scale=1.0]{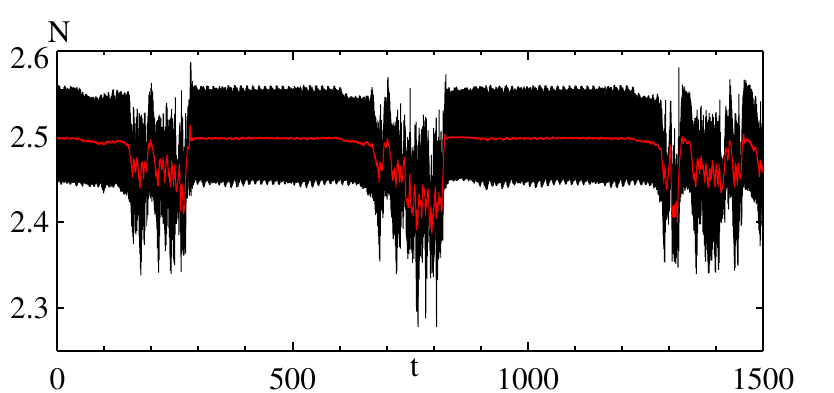}}
\vspace*{-0.5cm}
\centerline{\hspace*{-8.2cm}(b)}
\centerline{\includegraphics[scale=1.0]{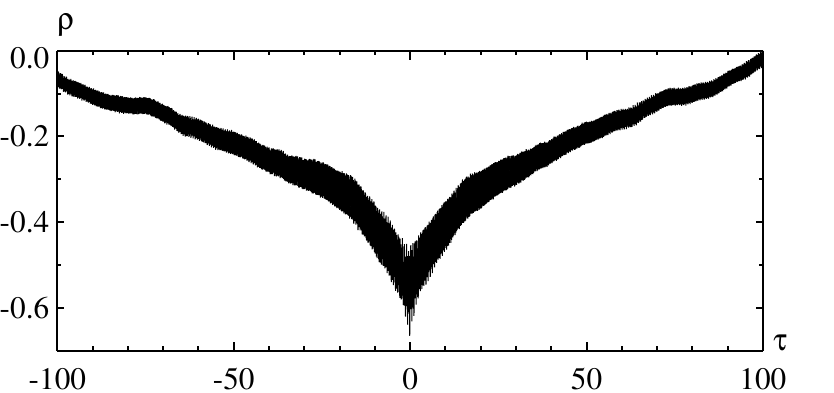}}
\vspace*{-0.5cm}
\centerline{\hspace*{-8.4cm}(c)}
\caption{(Color online)
(a): Evolution of the kinetic, $E_v(t)$, (black line, left axis) and logarithm of magnetic, $E_b(t)$,
(red line, right axis) energies for the convective attractor at Ra=2130 and $P_m~=~3$.
Thick blue lines in the horizontal axis represent time intervals, 
where the trajectory in the phase space is in transition before recovering the convective attractor.
Magnetic energy evolution for the same Ra and $P_m=2$ is represented by the red dashed line.
(b): Evolution of the spectral average, $N(t)$, (black line) and its forward moving average over 
time interval of length 2 (red line) for the regime at Ra=2130 and $P_m=3$ shown in (a).
(c): Cross correlation, $\rho(\tau)$, of the kinetic energy shown in (a) and the 
spectral average shown in (b); $\tau$ stands for lag measured in units of time.
}
\label{fig:ra2130pm3}
\end{figure}

The critical magnetic Prandtl number, $P_m^c$, for the chaotic convective
attractor at Ra=2070 is $9<P_m^c<10$ (see fig.~\ref{fig:ra2070pm10}(a)), while for
the regular state at Ra=2130, it is $2<P_m^c<3$ (see fig.~\ref{fig:ra2130pm3}(a)). 
This significant difference in $P_m^c$ is crucial for the onset of magnetic field 
generation by the convective attractors for larger values of Ra, 
where both regular and bursty phases are destabilized, but visited intermittently.

At Ra=2070, for the supercritical value $P_m=10$, after an initial exponential growth
the magnetic field generated by the convective attractor reaches a chaotic state with
a relatively small fluctuation (see fig.~\ref{fig:ra2070pm10}(a), upper red line). 
In this transition to the saturated MHD state,  
the kinetic energy undergoes a significant decrease (black line in fig.~\ref{fig:ra2070pm10}(a)).
In the saturated state, a dominant spatial feature of magnetic fields 
is its concentration in half-ropes located near the horizontal boundaries 
(see a snapshot of isosurfaces of magnetic energy density, $({\bf b}\cdot {\bf b})/2$, 
in fig.~\ref{fig:ra2070pm10}(b)).  
This feature is common for all magnetohydrodynamic regimes found in the paper; 
such configuration of magnetic field was observed in many convective dynamo simulations with 
perfectly electrically conducting boundaries (see, e.g., \cite{physd,stpier93}).

The magnetic field generated by the regular convective attractor at Ra=2130 and $P_m=3$ is 
represented in fig.~\ref{fig:ra2130pm3}(a), and shows a very different behaviour.
After the initial exponential growth ($0\le t\le50$), the magnetic field saturates 
at $0.3\le E_b(t)\le 1.6$ for $50\le t\le 160$. Affected by the stronger magnetic field,
the perturbed convective attractor loses its previously laminar behaviour and undergoes a chaotic burst.  
During the burst, it ceases to generate 
magnetic field and for $160\le t\le 290$ 
the magnetic energy decreases by 42 orders of magnitude. 
At $t\approx280$ the magnetic energy is low enough so that the unperturbed regular
convective attractor is recovered and the magnetic field gets reamplified by dynamo action. 
This scenario repeats intermittently in time.
The time to recover the convective attractor from the magnetic perturbation
varies significantly: in fig.~\ref{fig:ra2130pm3}(a) it is 250 time units (from $t=50$ to $t=300$)
and 310 (from $t=600$ to $t=910$) (these time intervals are illustrated by thick blue lines 
in the horizontal axis). An interesting feature of this magnetohydrodynamic regime
is that the magnetic field is generated by a less energetic and more regular (in time) phase. 

This intermittent switching between 
the amagnetic (hydrodynamic) and the magnetic states can be named as a 
``self-killing-and-self-recreating'' dynamo, in contrast to the ``self-killing'' 
dynamos \cite{fuchs99}, where the generated magnetic field modifies 
the flow in such a way that it is attracted to a non-generating stable hydrodynamic state.
Similar, although periodic in time, switching between a generating steady hydrodynamic state
and an unstable steady MHD state was found in \cite{physd} 
(see fig.~22 {\it ibid}.), where magnetic field generation in the rotating 
RBC was studied at various rotation rates. 

\begin{figure}[!]
\centerline{\includegraphics[scale=1.0]{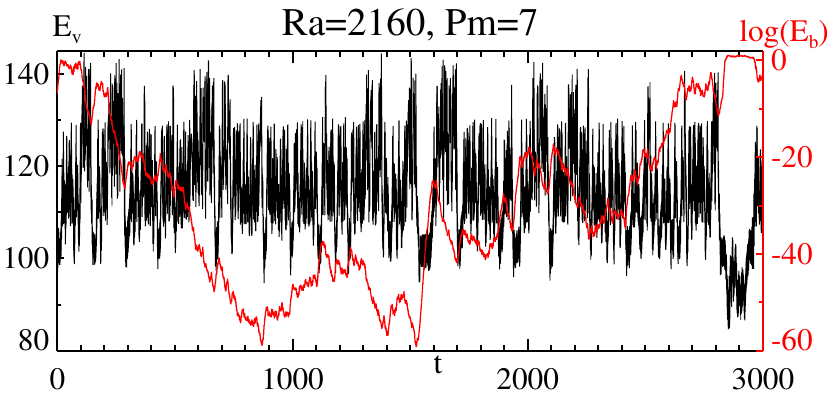}}
\vspace*{-0.5cm}
\centerline{\hspace*{-8.2cm}(a)}
\vspace*{0.00cm}
\centerline{\includegraphics[scale=1.0]{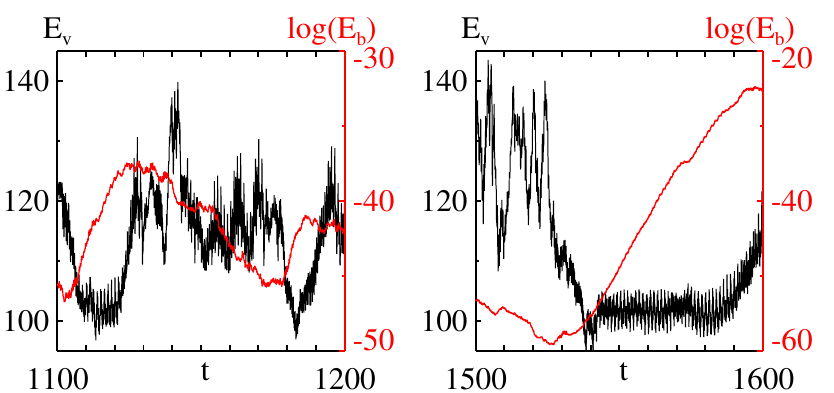}}
\vspace*{-0.5cm}
\centerline{\hspace*{-4.1cm}(b)\hspace*{3.8cm}(c)}
\caption{(Color online) 
Evolution of the kinetic (black line, left axis) and logarithm of magnetic 
(red line, right axis) energies for the convective attractor at Ra=2160.
Full time interval is shown in (a) and two time windows are represented in 
(b) and~(c).
}
\label{fig:ra2160pm67}
\end{figure}

The intermittent convective attractor for Ra=2160 (see fig.~\ref{fig:conven}(b)) 
does not generate magnetic field for $P_m\le 6$. For $P_m=7$, the generated magnetic 
field inherits the intermittent behaviour from the flow -- when the regime is near the 
regular state, the magnetic field is amplified (see fig.~\ref{fig:ra2160pm67} (b) in 
$1105\le t\le 1130$ and (c) in $1530\le t\le 1600$);
when the state is near the bursty phases, 
the magnetic field is decaying 
(see fig.~\ref{fig:ra2160pm67} (b) in $1130\le t\le 1170$). 
In contrast to the case at Ra=2130, here the origin of the transition from 
the regular to the bursty state is not the influence of the generated magnetic field, 
but the intrinsic intermittency due to the instability of both states in the purely hydrodynamic regime.

\begin{figure}[!]
\centerline{
\includegraphics[scale=1.0]{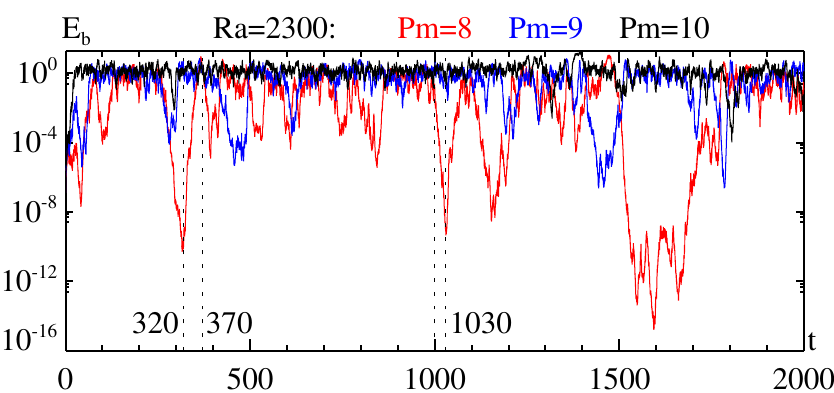}
\vspace*{-0.2cm}
}
\caption{(Color online)
Evolution of the magnetic energy generated by the convective attractor 
at Ra=2300 and for $P_m=8$ (red), $P_m=9$ (blue) and $P_m=10$ (black).
}
\label{fig:ra2300eh}
\end{figure}

Figure~\ref{fig:ra2300eh} shows the time series of the magnetic energy at Ra=2300
for different values of ${P_m}$. For the slightly supercritical value ${P_m=8}$ (red line), 
it displays intermittency, with an interplay of the
regular states amplifying the magnetic field (e.g. for $320\le t\le370$) 
and bursty states suppressing its generation (e.g. for $1000\le t\le1030$).  
The amplitude of magnetic energy variation is 16 orders of magnitude. The generated magnetic
field is weak, as well as its influence on the flow, so the dynamo is subordinated
to the switching between the regular and irregular phases of the convective
regime.  On increasing $P_m$ to 9 (blue line), and then 10 (black line), 
the fluctuations of the magnetic energy become smaller.
Since larger values of $P_m$ with fixed $P$ correspond to weaker magnetic
diffusion, the stronger magnetic fields bring the resulting
MHD regimes farther from the hydrodynamic one.  

The same estimate
of $P_m^c$ and the same behaviour of the generated magnetic field near onset
extends to the convective attractors at Ra=2400 and 2500. 

\begin{figure}
\centerline{
\includegraphics[scale=1.0]{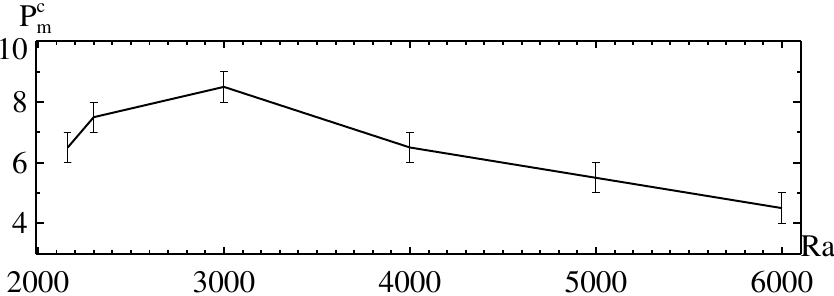}
}
\caption{Dependence of the critical magnetic Prandtl number estimate on Ra for the 
hyperchaotic regimes.
}
\label{fig:pmcra}
\end{figure}

For the convective attractor at Ra=3000 the critical magnetic Prandtl number
achieves its maximal value along the considered hyperchaotic convective regimes,
${8<P_m^c<9}$.  On further increase of Ra, for the sequence of convective attractors
at ${\rm Ra}=3000, 4000, 5000$ and 6000, where the regular phases are no longer seen,
the critical magnetic Prandtl number decreases with Ra (see table~\ref{tab:pmc} 
and fig.~\ref{fig:pmcra}). 
Simulations with the doubled resolution 
confirm this tendency for much higher values of Ra.
This is in accordance with findings referred 
in Busse \cite{busse00} for large values of Ra.

In order to understand why the quiescent phases of the kinetic energy time series
correspond to better dynamos, we characterized the hydrodynamic states by a measure
of the number of active spatial Fourier modes. The spectral average has been frequently employed in
this context \cite{rempel09}, being defined as 
$$
N(t)=\sqrt{ 
\sum_{n=1}^\infty \sum_{{\bf k}\in C_n} |{\bf k}|^2 |\hat{\bf v}_{\bf k}(t)|^2   /
\sum_{n=1}^\infty \sum_{{\bf k}\in C_n} |\hat{\bf v}_{\bf k}(t)|^2 
}  
$$
where $\hat{\bf v}_{\bf k}$ denotes the $k$th Fourier coefficient of the velocity field, and 
$C_n=\{ {\bf k}\colon n-1<|{\bf k}|\le n \}$ stands for the $n$th ($n\in\mathbb{N}$) 
spherical shell in the space of Fourier wave vectors ${\bf k}$.
Note that the spectral average is the square root of the averaged $|{\bf k}|^2$, where the average is weighted
by the shell-integrated energy. Therefore, it measures the energy
spread in the ${\bf k}$ spectrum, and should increase with time in systems
with energy cascade until dissipative effects restrain its growth. 
It can also be seen as the square-root of the ratio of the enstrophy to the energy 
and, consequently, as the inverse of a length scale related to viscous dissipation.
We employ the spectral average $N(t)$ as a 
measure of the effective number of degrees of freedom in the convective system. 

\begin{figure}
\centerline{\includegraphics[scale=1.0]{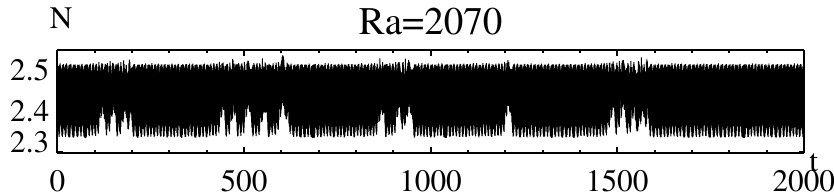}}
\vspace*{-0.5cm}
\centerline{\hspace*{-8.2cm}(a)}
\vspace*{0.00cm}
\centerline{\includegraphics[scale=1.0]{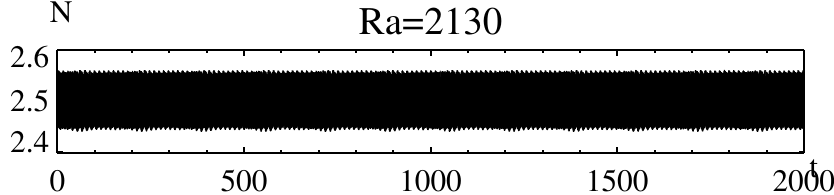}}
\vspace*{-0.5cm}
\centerline{\hspace*{-8.2cm}(b)}
\vspace*{0.00cm}
\centerline{\includegraphics[scale=1.0]{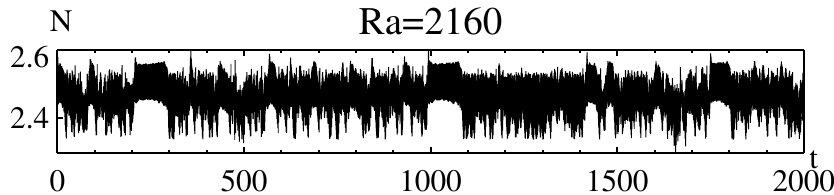}}
\vspace*{-0.5cm}
\centerline{\hspace*{-8.2cm}(c)}
\vspace*{0.00cm}
\centerline{\includegraphics[scale=1.0]{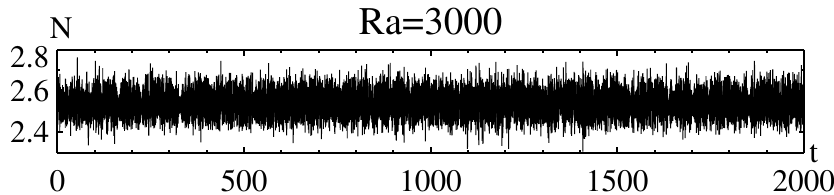}}
\vspace*{-0.5cm}
\centerline{\hspace*{-8.2cm}(d)}
\vspace*{0.00cm}
\centerline{\includegraphics[scale=1.0]{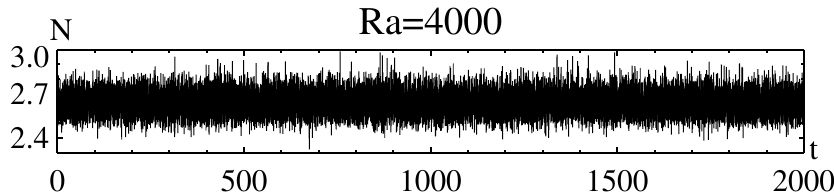}}
\vspace*{-0.5cm}
\centerline{\hspace*{-8.2cm}(e)}
\vspace*{0.00cm}
\centerline{\includegraphics[scale=1.0]{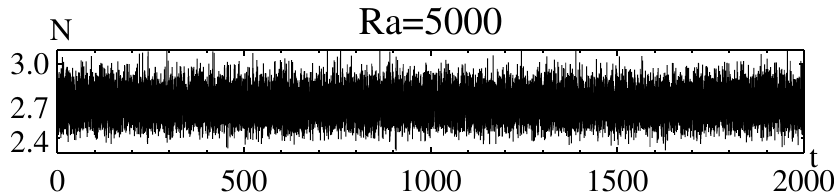}}
\vspace*{-0.5cm}
\centerline{\hspace*{-8.2cm}(f)}
\caption{Time series of the spectral average $N(t)$ for the convective attractors in the absence of magnetic field 
for Ra=2070~(a), 2130 (b), 2160 (c), 3000 (d), 4000 (e) and 5000 (f).}
\label{fig:medida}
\end{figure}

Fig. \ref{fig:medida}
plots the time series of $N(t)$ for some of the hydrodynamic regimes of fig. \ref{fig:conven}.
Note that the chaotic attractor at Ra=2070 has lower average $N(t)$ than the quasiperiodic
attractor at Ra=2130; consequently, the regular phases in the intermittent series at
Ra=2160 display higher $N(t)$ than the bursty phases. The higher values of $N(t)$ found
in the quiescent phases correlates well with the periods of magnetic field growth 
(for the same regime at Ra=2130 and $P_m=3$ cf. evolution of the kinetic and magnetic energies 
in fig.~\ref{fig:ra2130pm3}(a) with the spectral average, $N(t)$, in (b); in order to demonstrate
the negative correlation of the kinetic energy and the spectral average the cross correlation function of 
these quantities, $\rho(\tau)$, is shown in (c)).
For higher values of Ra (two lower panels of fig.~\ref{fig:medida}), 
the system dynamics is strongly irregular and the average $N(t)$ becomes larger, 
which coincides with the monotonic decay of the critical $P_m$ for dynamo action in these cases.

\section{Conclusions}

We have shown how intermittent convective attractors with two or more qualitatively
different (quiescent and bursty) phases 
can lead to intermittent dynamo action.
The critical $P_m$ for dynamo action is lower for the quasiperiodic/quiescent attractors
compared to the hyperchaotic/bursty attractors. As Ra is increased, the quiescent phases in
the intermittent time series are shortened
and, thus, the overall critical $P_m$ also becomes larger due to the predominance of bursts
in the time series. Further increase in Ra leads the convective system to a state where
quiescent phases are no longer observable and, then, the critical $P_m$ starts to decrease
with increasing Ra, as expected in strongly hyperchaotic/turbulent systems.  Therefore, in
transition to turbulence, quasiperiodic states are preferred for dynamo action in the
intermittent regimes, but in stronger chaotic regimes without this type of chaotic
intermittency, the critical $P_m$ is expected to become lower and lower as Ra is increased.
This low-$P_m$ MHD regime  is  of  particular  interest, since many geophysical
and astrophysical problems are described by low $P_m$. For example, in the Sun, $P_m$ varies
between $10^{-7}$ and $10^{-4}$ between the top and the bottom of the convection zone
\cite{axel_chian}, whereas in the geodynamo it is approximately $5\times 10^{-6}$
\cite{kono02}. 

The intermittent switching between qualitatively different phases in the time series of
kinetic energy shown in figs. \ref{fig:conven} and \ref{fig:ra2130pm3}(a) resembles the
bi-stability reported by Zimmerman {\it et al}. \cite{zimm11} in experiments with a rotating
spherical Couette flow and later associated with the intermittent behavior of the magnetic
field in an MHD experiment with an imposed magnetic field \cite{zimm14}. 
For instance, fig. 10 of ref. \cite{zimm14}
shows time series of torque and azimuthal magnetic field where the system undergoes
intermittent transitions between two phases with very different mean values.  The
intermittency in our magnetic energy time series also shows similarity with the on-off
intermittent ABC-flow dynamo reported by Rempel {\it et al}. \cite{rempel09}, where the
magnetic energy time series intercalate bursty phases with quiescent phases of almost null
magnetic energy.  This type of intermittency is expected to happen near critical values of
the control parameters where a global bifurcation in the underlying attractor takes place
\cite{sweet01,ponty}, and has recently been found in numerical simulations of rotating
spherical Couette flows with realistic boundary conditions \cite{raynaud13}.

The verification that the quasiperiodic quiescent phases of the velocity field constitute a
better dynamical state for a convective dynamo than the bursty phases for low Ra may seem at
odds with the intuitive idea that a chaotic flow should favor the stretching, twisting and
folding (STF) of magnetic field lines, which is a well known mechanism for magnetic field
amplification \cite{fastdynamo}. However, one should note that the STF dynamo may operate
even in stationary flows with chaotic streamlines. 
Inhibition of the large-scale magnetic field by large fluctuations of the velocity field  
is shown analytically in \cite{pe06} for some simple flows using the mean-field dynamo theory.
In our case, we have shown that the
quiescent/regular phases in the intermittent regime excite a higher number of spatial
Fourier modes than the bursty phases, which implies the increase in magnetic flux in the
regular phases. For higher values of Ra, when the energy cascade and spatiotemporal complexity
increase, the critical $P_m$ decreases. It would be interesting to check if the same
happens in spherical geometries, in a set up more closely related to the difficult task of
magnetic field generation in laboratory experiments.  As a final remark, it is worth
mentioning that the behavior of the dynamo can be quite different in the presence of
rotation and that should be the topic of future exploration.

\section*{Acknowledgments}
RC, ELR and EVC acknowledge financial support from FAPESP
(2013/01242-8, 2013/26258-4 and 2016/07398-8, respectively). 
RC was partially financed by the project POCI--01--0145--FEDER--006933 SYSTEC 
funded by FEDER funds through COMPETE 2020 and by FCT (Portugal). 
ELR also acknowledges financial support from CNPq (grant 305540/2014-9)
and CAPES (grant 88881.068051/2014-01).

\bibliographystyle{elsarticle-num}
\bibliography{pap}

\end{document}